\documentclass[11pt,a4paper]{article}
\usepackage[utf8]{inputenc}
\usepackage[T1]{fontenc}
\usepackage{graphicx}
\usepackage{authblk}

\title{The role of haplotype complementation and purifying selection in the genome evolution}

\author[1]{Stanis\l{}aw Cebrat}
\author[2]{Dietrich Stauffer}
\author[1]{Wojciech Waga}

\affil[1]{Department of Genomics, Faculty of Biotechnology, University of Wroc{\l}aw, ul. Przybyszewskiego 63/77, 51-148 Wroc{\l}aw, Poland. e-mail: cebrat@smorfland.uni.wroc.pl}
\affil[2]{Institute for Theoretical Physics, Cologne University, D-50923 Köln, Euroland}

\begin{document}
\maketitle

\begin{abstract}
We discuss two different ways of chromosomes' and genomes' evolution. Purifying selection dominates in large panmictic populations, where Mendelian law of independent gene assortment is valid. If the populations are small, recombination processes are not effective enough to ensure an independent assortment of linked genes and larger clusters of genes could be inherited as the genetic units. There are whole clusters of genes which tend to complement in such conditions instead of single pairs of alleles like in the case of purifying selection. Computer simulations have shown that switching in-between complementation and purification strategies has a character of a phase transition. This is also responsible for specific distribution of recombination events observed along eukaryotic chromosomes - higher recombination rate is observed in subtelomeric regions than in central parts of chromosomes - for sympatric speciation and probably for non-monotonous relation between reproduction potential and genetic distance between parents.
\end{abstract}

\section*{Key words:} sympatric speciation, age structured populations, genetic complementation, evolution, Monte Carlo simulations.

\section{Introduction}
\begin{figure}[h]
\centering
\includegraphics[width=9cm]{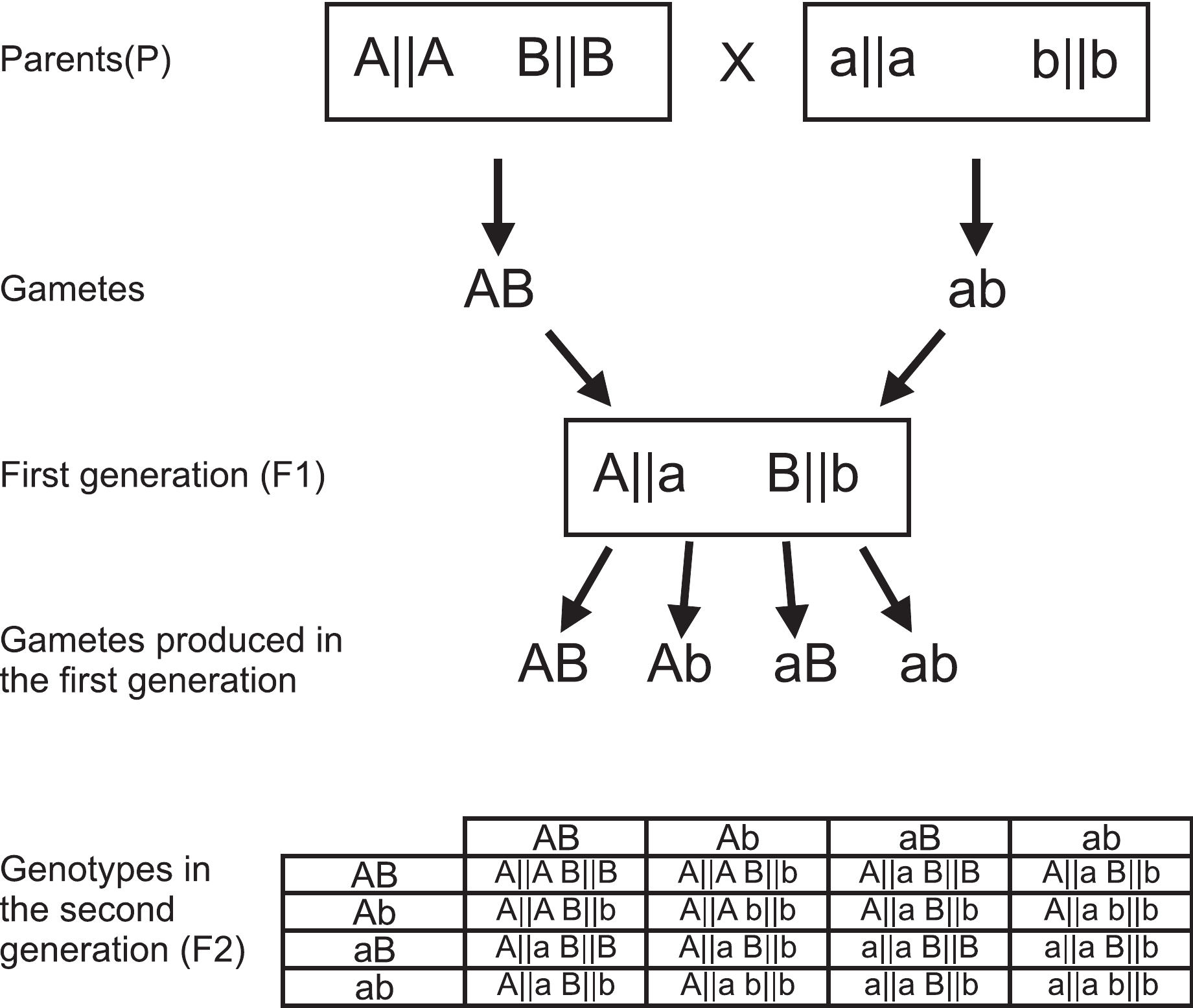}
\caption{Scheme of inheritance of two pairs of alleles in case when they assort into gametes independently and quantitative relations between genotypes in the second generation.}
\label{f1}
\end{figure}

Contemporary, common knowledge of genetics is based mainly on Mendelian laws of inheritance integrated with Morgan chromosomal theory. Mendel concluded, from the results of his experiments, that there should be some hereditary units (he called them ''factors'') which are responsible for transferring some traits from parents to their offspring. These factors can be dominant, like that one responsible for purple colour of pea flowers or recessive, like that one responsible for white flowers of pea. The last one is hidden in the plants which possess both types of factors. In the normal somatic cell (diploid) two corresponding factors exist for each trait but they are separated when a haploid gamete is produced during meiosis. Mendel's laws describe such a separation of factors belonging to one pair responsible for a given trait or belonging to more pairs responsible for different traits.
The first Mendel law - Law of Segregation or purity of gametes - says that each gamete receives only one factor of a pair (one allele of a gene). The second Mendel law - Law of Independent Assortment - says that alleles of different genes assort independently of each other during gamete formation. The Mendel Laws were described in 1865 \cite{Mendel} and rediscovered by de Vries in 1900 \cite{vries}.  A few years later they have been supported by Boveri \cite{boveri} and Sutton \cite{Sutton} who suggested that chromosomes can carry the hereditary traits intimated by Mendel. But it was T. Morgan who, after many experiments performed with \emph{D. melanogaster}, supplied the clear evidence that the chromosomes are the real vehicles of hereditary information. He published with his colleagues in 1915 a paper \emph{Mechanism of Mendelian Heredity} \cite{Morgan}. They stated that Mendelian factors are like beads located on chromosomes - pairs of the corresponding beads on the pairs of homologous chromosomes. They segregate with chromosomes into germ cells and again combine in zygotes, exactly as predicted by Mendel. They noticed that some traits are sex linked which means that these traits are determined by ''beads'' located on the chromosome which determines gender. Furthermore, they noticed that some traits can be linked and tend to transfer together to the gametes just because they probably are located close to each other on the same chromosome. But sometimes those connections are broken and they can assort to the gametes independently producing a new combination of traits. It is the last notice which triggers the common conviction that all alleles are finally assorted independently into the gametes. In fact, it could be true only in very large populations sharing the common genetic pool, called Mendelian population. It is not a case in natural populations with spatially distributed individuals, where genetic relations (or distances) between individuals have to be taken under account.
Let's consider a situation presented in Fig.\ref{f1}, describing the classic Mendel's laws. Each parental individual ($P$) possesses two pairs of alleles located on different chromosomes, the first parent has all alleles dominant: $A||A$, $B||B$ and the second one - all alleles recessive: $a||a$ and $b||b$. Since each of these individuals can produce only one type of gametes: $AB$ or $ab$, in the first generation ($F1$) all individuals have genotypes $A||a$ and $B||b$ and phenotypes determined by the dominant alleles. Each of these $F1$ individuals can produce four types of gametes: $AB$, $Ab$, $aB$, and $ab$. Assuming independent and random assortment of chromosomes, the fraction of each type of gametes is $0.25$. That determines the ratios between genotypes of the $F2$ generation, assuming no gamete preselection. The genotypes shown in the table determine the phenotypes of individuals with ratios: $9/16$ individuals with both traits determined by dominant alleles, $3/16$ for each of two phenotypes with one dominant and recessive trait and $1/16$ for the phenotype with both recessive traits.

\begin{figure}
\centering
\includegraphics[angle=0,width=9cm]{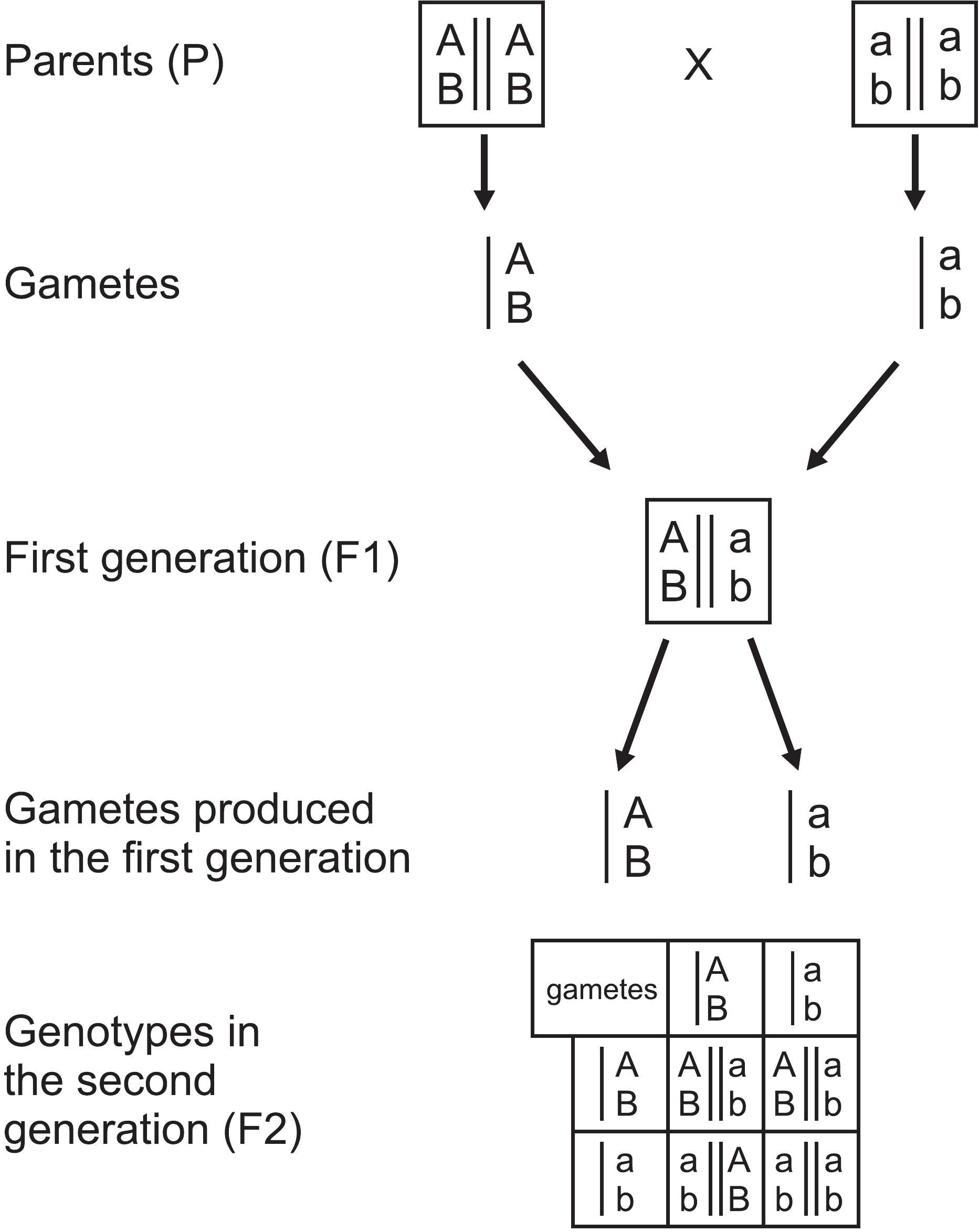}
\caption{Scheme of inheritance of two pairs of alleles in case when they are linked and assort into gametes as a cluster and quantitative relations between genotypes in the second generation.}
\label{f2}
\end{figure}

Now, let's analyze the situation where the considered pair of genes is located on the same chromosome close to each other: $AB||AB$ and $ab||ab$. The parental individuals can produce gametes with the same genetic contents as in the previous case. All $F1$ individuals are also $AB||ab$. Nevertheless these $F1$ individuals are going to produce only $AB$ and $ab$ gametes and the $F2$ generation will be much less genetically and phenotypically diversified, only $AB||AB$, $AB||ab$, $aa||AB$ and $ab||ab$ with a corresponding phenotypic ratio: $3:1$.
Let's assume that $A$ and $B$ denote functional alleles of genes responsible for important functions of an organisms while a and b are their corresponding non-functional, deleterious alleles. Any organism with $a||a$ and/or $b||b$ genetic configuration has to suffer from genetic disease(s). It is obvious that the two above situations are quite different. In the case of independent assortment Fig.\ref{f1} the fraction of $7/16$ of $F2$ would be suffering while in the second case Fig.\ref{f2} only $1/4$ is going to suffer. It is obvious that selection would favour some genetic configurations. Paradoxically, there are no recombinations between alleles in the second case though recombinations suppose to ameliorate the genetic conditions of genomes. 

\begin{figure}
\centering
\includegraphics[angle=0,width=9cm]{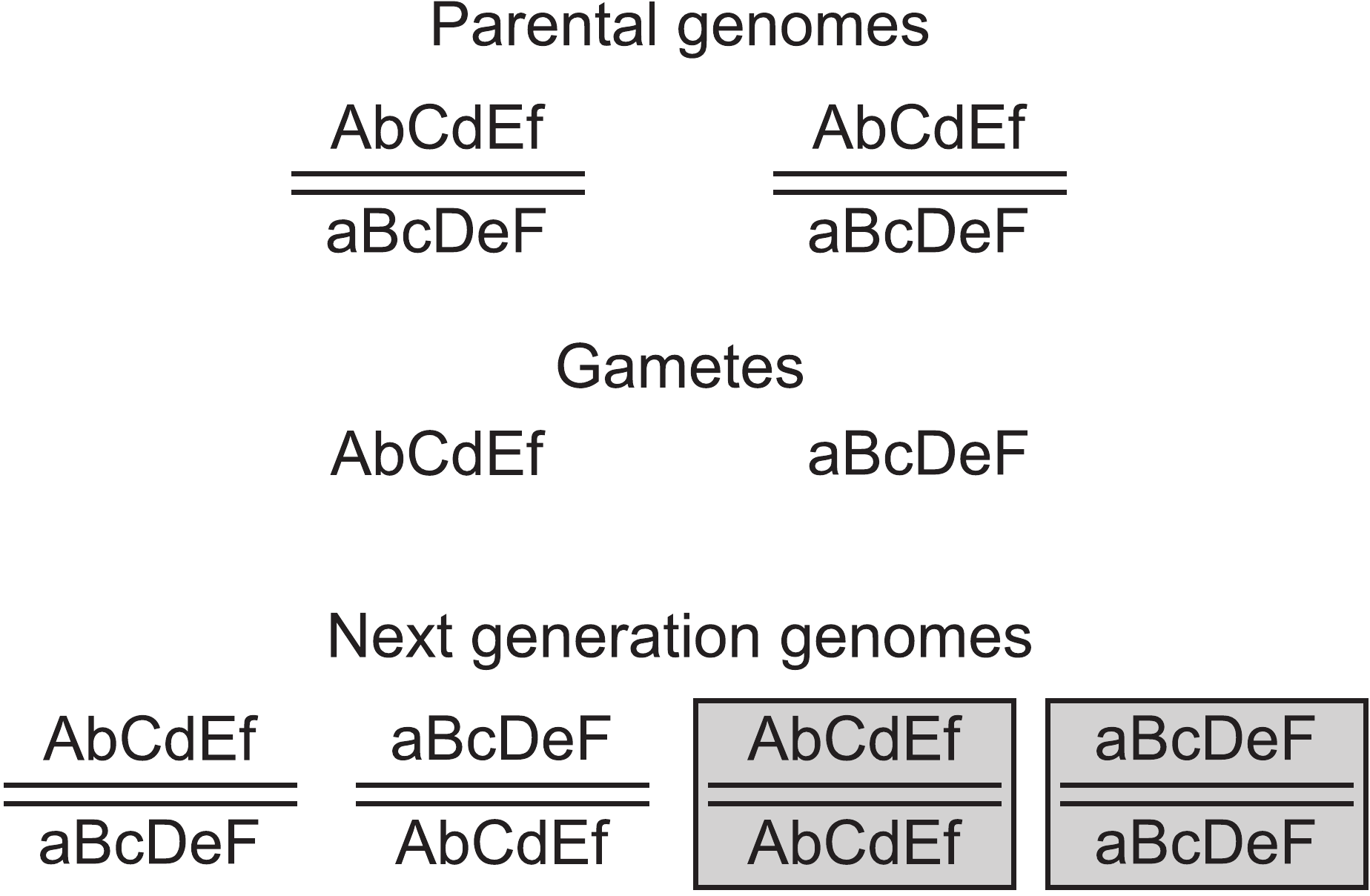}
\caption{Scheme of inheritance of complementing gene cluster. Notice that without recombination, if gametes fuse randomly, half of individuals in the next generation possess complementing haplotypes while half of them have pairs of identical haplotypes, they express many deleterious phenotypic traits (grey boxes).}
\label{f3}
\end{figure}

\begin{figure}
\centering
\includegraphics[angle=0,width=9cm]{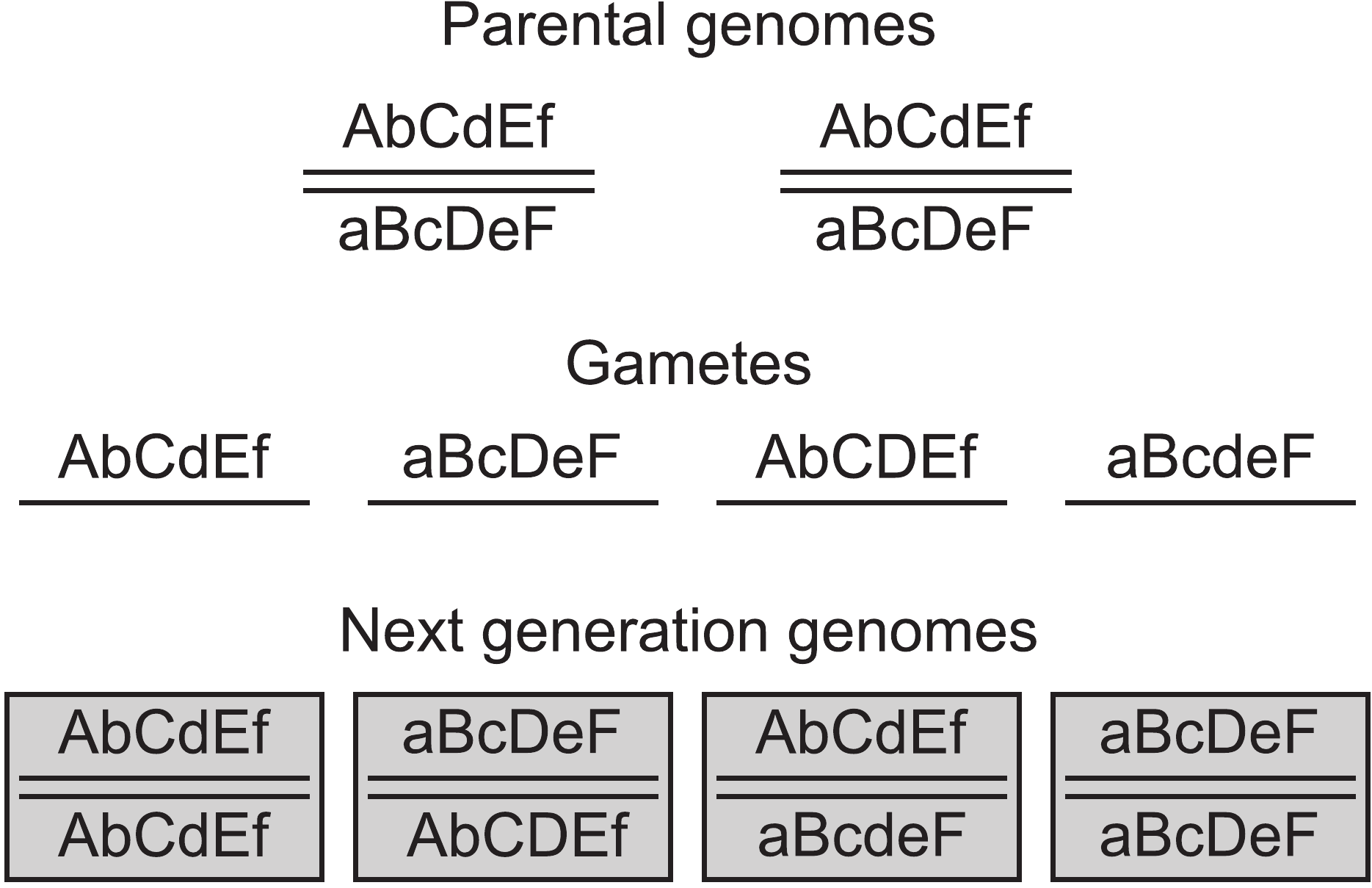}
\caption{Scheme of inheritance of complementing gene cluster with one recombination event during gamete production in one parental genome. Notice that recombination in the second (left) parent produces gametes which can not form fully complementing zygotes with non-recombined gametes. Thus, all individuals in the next generation possess pairs of recessive alleles and express many deleterious phenotypic traits (all boxes are grey).}
\label{f4}
\end{figure}

A situation where more linked and complementing alleles occur in the heterozygous state in the parental genomes is shown in Fig.\ref{f3}. ''Complementing'' means that dominant alleles hide (complement) the effect of their potentially deleterious, recessive counterparts. All considered loci in the parental genomes are occupied by pairs of different alleles - one recessive, another one dominant. Thus, the individuals have all phenotypic traits determined by the dominant alleles. Since in the described case we have assumed that all alleles are closely linked thus, there are no recombinations inside the group of genes (called cluster of genes), and only two types of gametes are produced. Assuming that these gametes combine randomly to form zygotes, half of the produced zygotes would be composed of complementing gametes while the other half would be composed of identical gametes (shadowed in the figure). In the last cases all loci would be homozygous and some of them would determine recessive traits. If there is a very low recombination probability inside the cluster, then we have a situation described in Fig.\ref{f4}. One individual produces the recombined gametes (the right two gametes). These gametes cannot ''find'' a fully complement gamete in the pool of non-recombined gametes produced by the rest of individuals in the population (the left two gametes). Thus, a gamete with the recombined cluster produces a zygote with some recessive characters which could die even before birth (zygotic deaths).

Let's stay with the example described in Fig.\ref{f4}. Both parental genomes have identical configurations of their haplotypes (clusters of genes) and recombination inside the cluster is low but possible so, it can happen with a finite frequency during each gamete production. What could happen with their genomes after many generations? If there is no selection for any specific genome configuration, mating is random, the number of generations is very high and population is very large, then one can expect that many recombinations can happen inside the clusters and the original allele configurations would be disrupted. In such a scenario alleles seems to be randomly assorted into gametes and the Hardy-Weinberg principle (HWP) is in force \cite{Hardy} \cite{wein08}.  The fractions of any allele combination can be calculated just from the frequencies of alleles in the genetic pool. Nevertheless, natural populations do not fulfil the conditions required for such a ''Mendelian population''. Especially, populations sharing a common genetic pool are usually much smaller than the whole population of the species, there are some mating preferences, subpopulations could be separated by physical or geographical barriers, selection favours or discriminates some genetic configurations and, what is the most important, all these conditions are superimposed on the restricted and unevenly distributed recombination events along the eukaryotic chromosomes. Thus, emerging of gene clusters should be possible and complementation of whole clusters in genomes should be also observable. In the next chapters we present computer models describing the conditions where complementation of large gene clusters is generated showing a possible way of eukaryotic chromosomes' evolution.

\section{Modelling the evolution of chromosomes}
\subsection{Populations without age structure}

Computer modelling of the population evolution were performed according to the Monte Carlo method described previously \cite{Zawierta} \cite{Zawierta2}. Virtual populations are composed of individuals represented by their diploid genomes. Genomes are composed of pairs of bitstrings (homologous chromosomes), each $L$ bits long. A bit set to $0$ corresponds to the wild (functional) form of an allele while a bit set to $1$ corresponds to the defective allele. It has been assumed that all defective alleles are recessive which means that two alleles at corresponding loci in both chromosomes (the same positions in bitstrings) have to be defective to determine deleterious phenotype. To reproduce, an individual seeks for a partner. Each one of two partners produces one gamete and the fusion of the gametes gives a newborn. Gametes are produced by replication of two parental bitstrings, introducing $M$ mutations into random positions of each new copy, and recombination between copies in the corresponding but random positions mimicking crossover. Mutation changes bits $0$ to $1$, if a bit chosen for mutation is already set to $1$ it stays one - there are no reversions. Frequency of mutations $M$ and recombinations $C$ are very important parameters of the model. After each Monte Carlo step (MCs) $2\%$ of individuals are randomly eliminated from the population. This gap is filled up in the next MCs by surviving newborns. Newborns with at least one position determining a deleterious phenotype (both bits at corresponding positions set to $1$) will never reproduce and are dying.

\subsection{Age structured populations}
To simulate the evolution of the age structured populations we have used diploid version of the Penna model \cite{Penna}. The Penna model differs from that one described above by consecutively expressing (switching on) the bits located on the bitstrings - in the first MCs after reproduction, a newborn expresses the first pair of bits, in the seconds MCs - the first two pairs and so on. Thus, the age of individuals corresponds to the number of switched on bits. If the maximum number of the expressed deleterious phenotype traits (threshold $T$) is reached, this individual is killed because of its genetic status. In such populations the maximum age equals $L$, but usually, after long enough simulations all individuals eventually die because of their genetic death before they reach the age $L$ and the age structures of simulated populations correspond very well to natural diploid populations. Additional parameters in this model are: minimum reproduction age $R$ - individuals which reach this age can reproduce, and $B$ - number of newborns which can be produced by one female during each MCs. The gender of each newborn is determined - female or male - with equal probability. In panmictic populations, a female at reproduction age looks for a male partner in the reproduction age in the whole population to produce an offspring which is born with probability described by Verhulst factor: $V=\frac{1-N_t}{N_{max}}$; where $N_t$ is an actual number of individuals in the population while $N_{max}$ is maximum size of population. There are no random deaths of individuals after birth \cite{Martins}. Each female being at reproduction age is chosen for reproduction only once during one MCs, while the male after reproduction is going back to the pool of males and again can be chosen by another female for reproduction. Such a population is called panmictic. To impose some restriction on panmictic reproduction it is possible to divide the whole population for smaller, non-overlapping subpopulations or to introduce a spatial distribution of individuals determining distances between sexual partners and their offspring, with partners selected from the neighbourhood and birth requiring a free nearby place. 

\subsection{Spatially distributed populations}
To study the role of spatial distribution of the large populations, individuals were distributed on a square lattice - at most one individual in one square. To avoid the border effect, the lattice was wrapped on a torus. To reproduce, the female chooses a partner located in a distance not larger than $P$ and puts an offspring at distance not larger than $O$ if there is any free square at such a distance. The role of spatial distribution was studied in both models described above - without age structure, introducing random elimination of declared fraction of population every MCs, or age structured populations, where individuals aged and were eliminated by genetic death only. 

\section{Results and discussion}

\subsection{Emerging complementarity; transition between complementing and purifying selection strategies}
\begin{figure}
\includegraphics[angle=0,width=14cm]{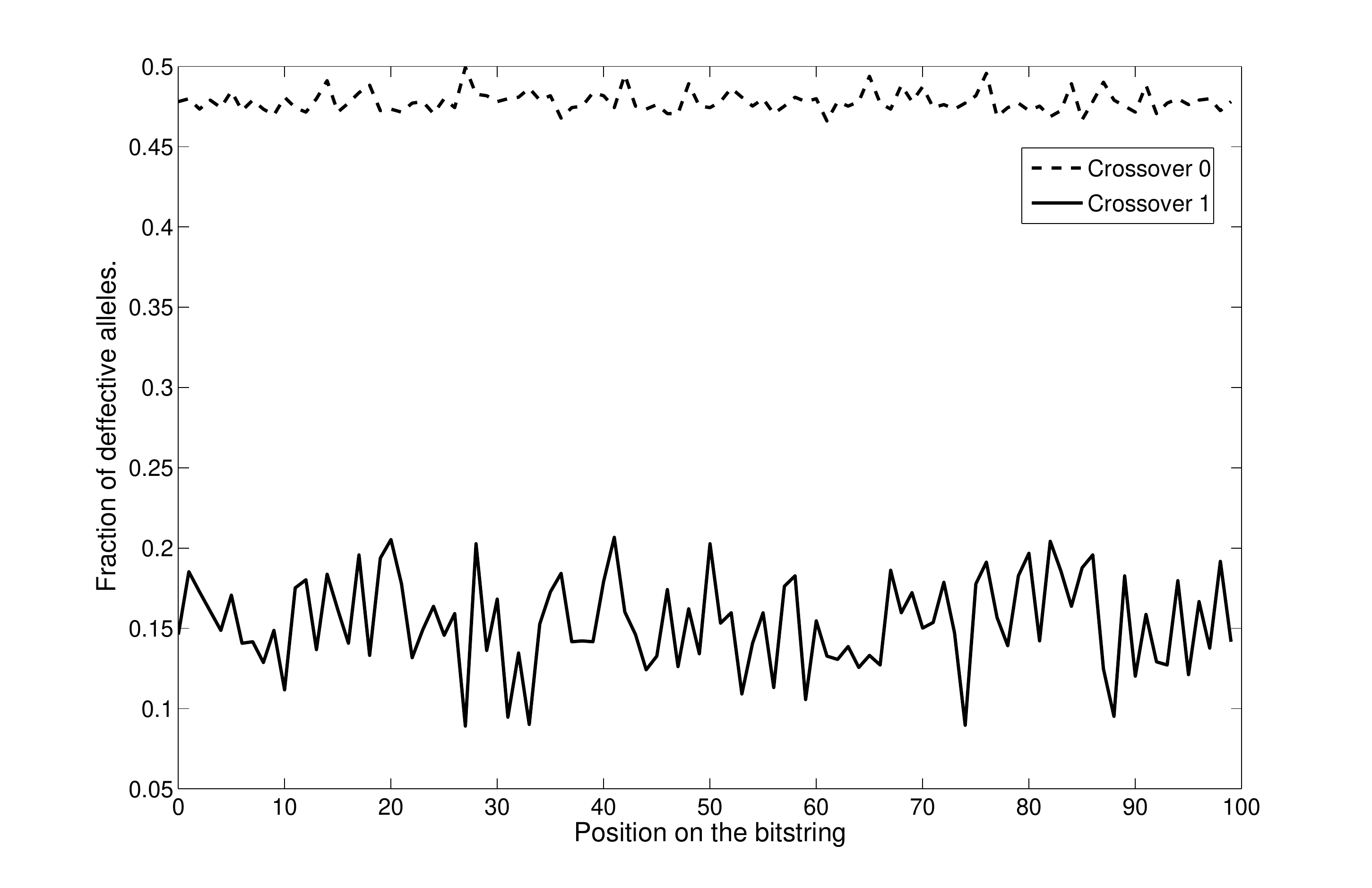}
\caption{Fractions of defective alleles in the diploid genomes with 1 crossover between homologous chromosomes and without recombinations. Number of alleles in one chromosome (bitstring) - 100.}
\label{f5}
\end{figure}

\begin{figure}
\includegraphics[angle=0,width=13cm]{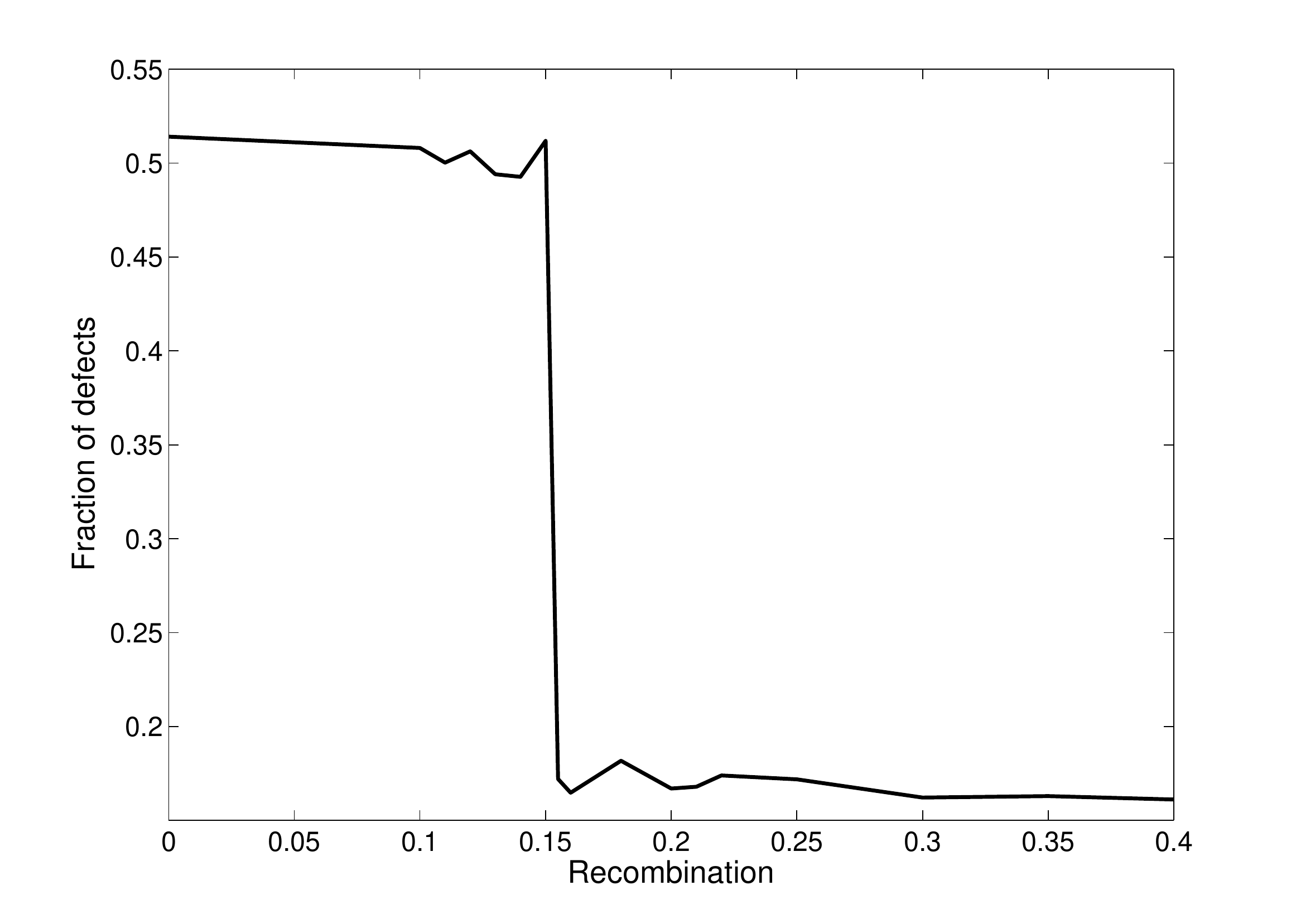}
\caption{Transition between complementation and purifying selection. Below recombination rate $0.14$ high fraction of defective alleles is observed and complementation prevails. At recombination rate $0.14-0.15$ (critical recom\
bination rate $C*$) sharp transition for purifying selection is observed.}
\label{f6}
\end{figure}

\begin{figure}
\includegraphics[angle=0,width=14cm]{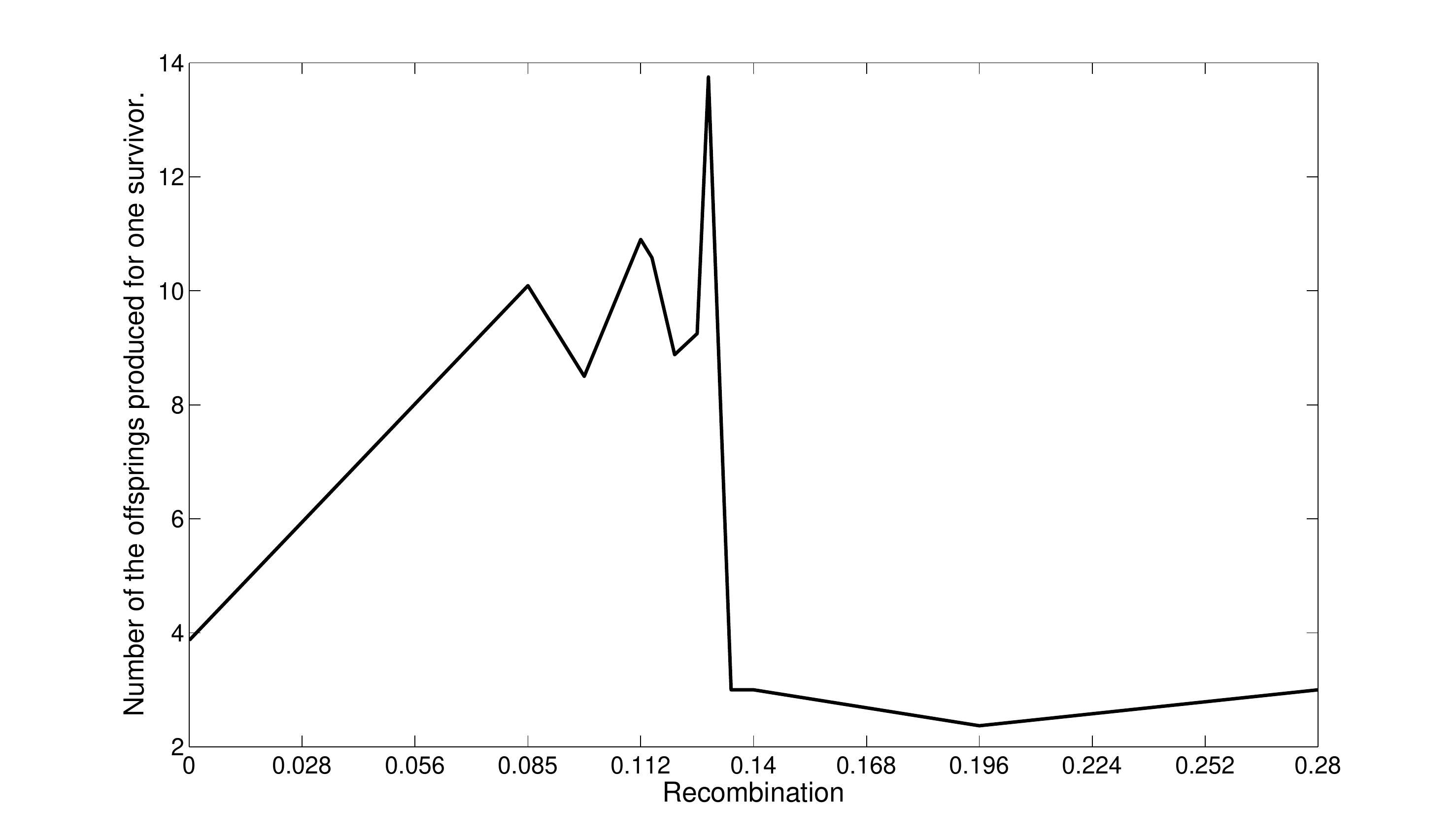}
\caption{Relation between the probability of forming a surviving zygote and recombination rate. Y-axis is scaled in the average number of offspring which have to be formed to get one survivor.}
\label{f7}
\end{figure}

Complementing groups of genes are described in the introduction and shown in Fig.\ref{f3}. The easiest way to show the emergence of complementation in computer simulations is to declare that there are no recombinations between bitstrings and that heterozygous loci increase the fitness of individuals, like in case of sickle cell anaemia trait \cite{Ragusa}, cystic fibrosis \cite{Gabriel} or Tay-Sacks syndrome in humans \cite{myerowitz}. Under such condition $(L=100, N=1000, M=1, C=0)$, in the first version of the model, genomes reach full heterozygosity (all loci are heterozygous) after about $10^5$ MCs \cite{Zawierta}. One recombination event between bitstrings during each gamete production prevents complementarity to emerge. In the fully heterozygous genome, the frequency of defective alleles is $0.5$. Since in the model, one homozygous locus with both alleles defective kills the individual, assuming that defective alleles are distributed randomly along the bitstring, the probability of forming a surviving, diploid genome without any locus with both defective alleles (homozygous) is of the order of $10^{-13}$. Populations under such conditions should be extinct, but they survive. Surviving is possible because the distribution of defects along the chromosomes is not random. Selection leads to the state where mainly two different types of bitstrings are present in the whole genetic pool. If two identical bitstrings form a newborn's genome - it is eliminated because many homozygous loci determining deleterious phenotypes appear in such a genome. If two different (complementing) bitstrings form a genome - it survives and it is the fittest because it has the maximum number of heterozygous loci. Thus, the probability of forming the surviving genome is not $10^{-13}$ but 0.5 in case when only two complementing haplotypes exist in the genetic pool. Recombinations disrupt the complementing haplotypes, as presented in Fig.\ref{f4}. Further simulations have shown that it is not necessary to declare the advantage of heterozygosity to produce complementing haplotypes (Fig.\ref{f5}). Simulations performed without recombinations between chromosomes during gamete production lead to full complementarity between haplotypes, while inclusion of recombinations between them prevents such a complementarity to emerge and purifying selection prevails. The question is - what is the character of the transition between the two strategies of the genome evolution under an intermediate recombination rate? Simulations performed under the same parameters: $L=100, M=1, N=1000$ with only one free parameter - recombination rate - have shown that populations choose between the two strategies around the recombination rate $0.14$ (Fig.\ref{f6}). For the above parameters, below this frequency of recombination, complementation prevails while above this value, the strategy of purifying selection is winning. The range of recombination rate where the transition between these strategies occurs is very narrow - of the order of $0.01$ \cite{Zawierta2}. The recombination rate at which the transition between the two strategies of evolution is observed has been called critical recombination rate – $C^*$.  At the critical recombination rate, the probability of forming a surviving zygote from two gametes is the lowest (see Fig.\ref{f7}).

\subsection{Relation between the critical recombination rate and population size}

\begin{figure}
\centering
\includegraphics[angle=0,width=\textwidth]{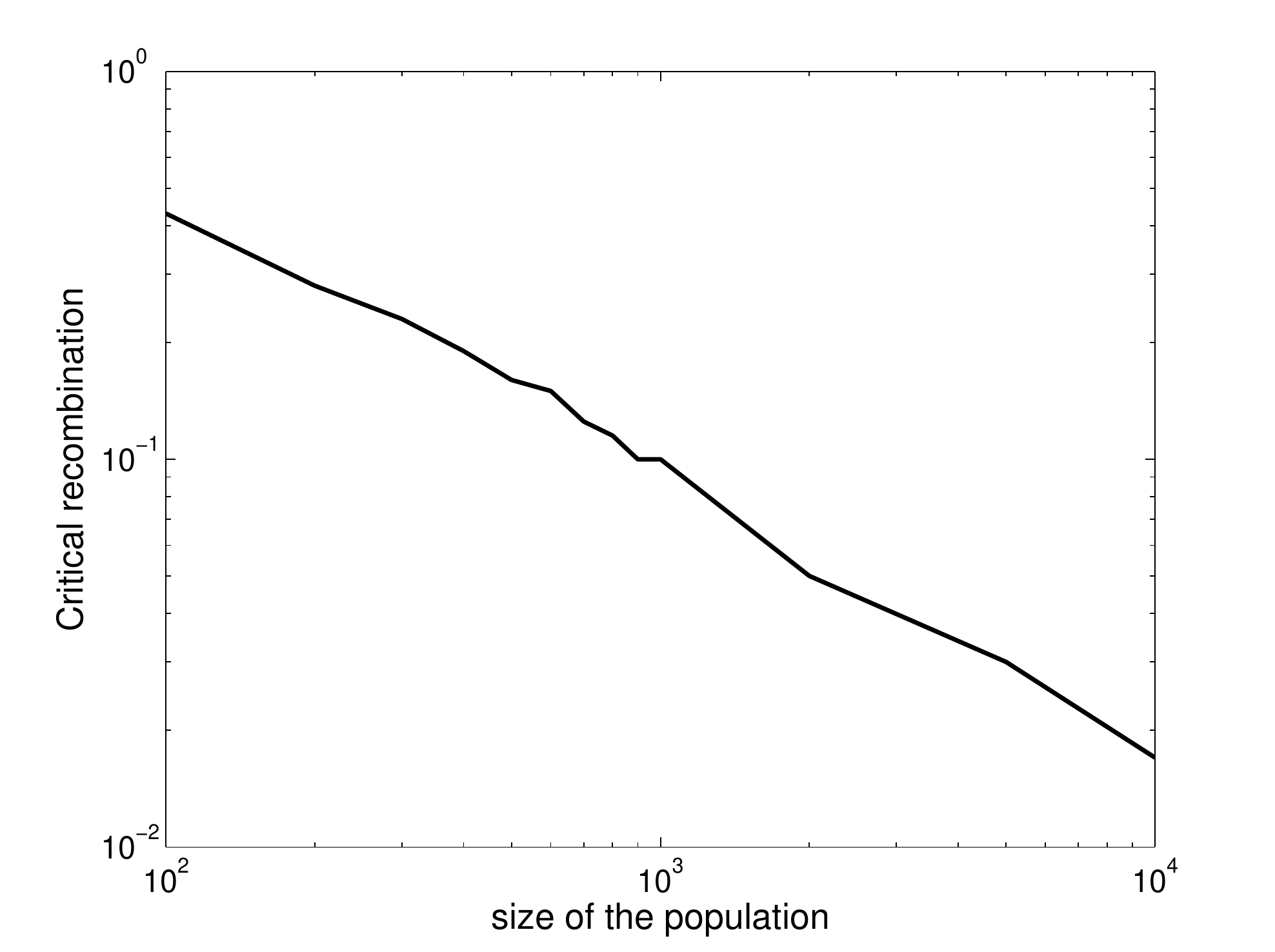}
\caption{Relation between critical recombination rate and effective population size. Notice the logarithmic scale of both axes.}
\label{f8}
\end{figure}

There is a power law relation between the value of critical recombination rate $C^*$ and the population size (Fig.\ref{f8}, notice the logarithmic scale of both axes). Larger panmictic populations have a lower critical value $C^*$ \cite{Zawierta2}. The explanation of this phenomenon seems to be trivial; in small populations the average genetic relation between individuals is high and the total number of crossover events in a given chromosome which could happen in the lineages of sexual partners to their common ancestor is relatively low - the inbreeding coefficient is high. In the larger populations this number is higher and the disruption of gene clusters is more probable. 

\subsection{Transition between complementarity and purifying selection strategies in spatially distributed populations}

\begin{figure}
\centering
\includegraphics[angle=0,width=\textwidth]{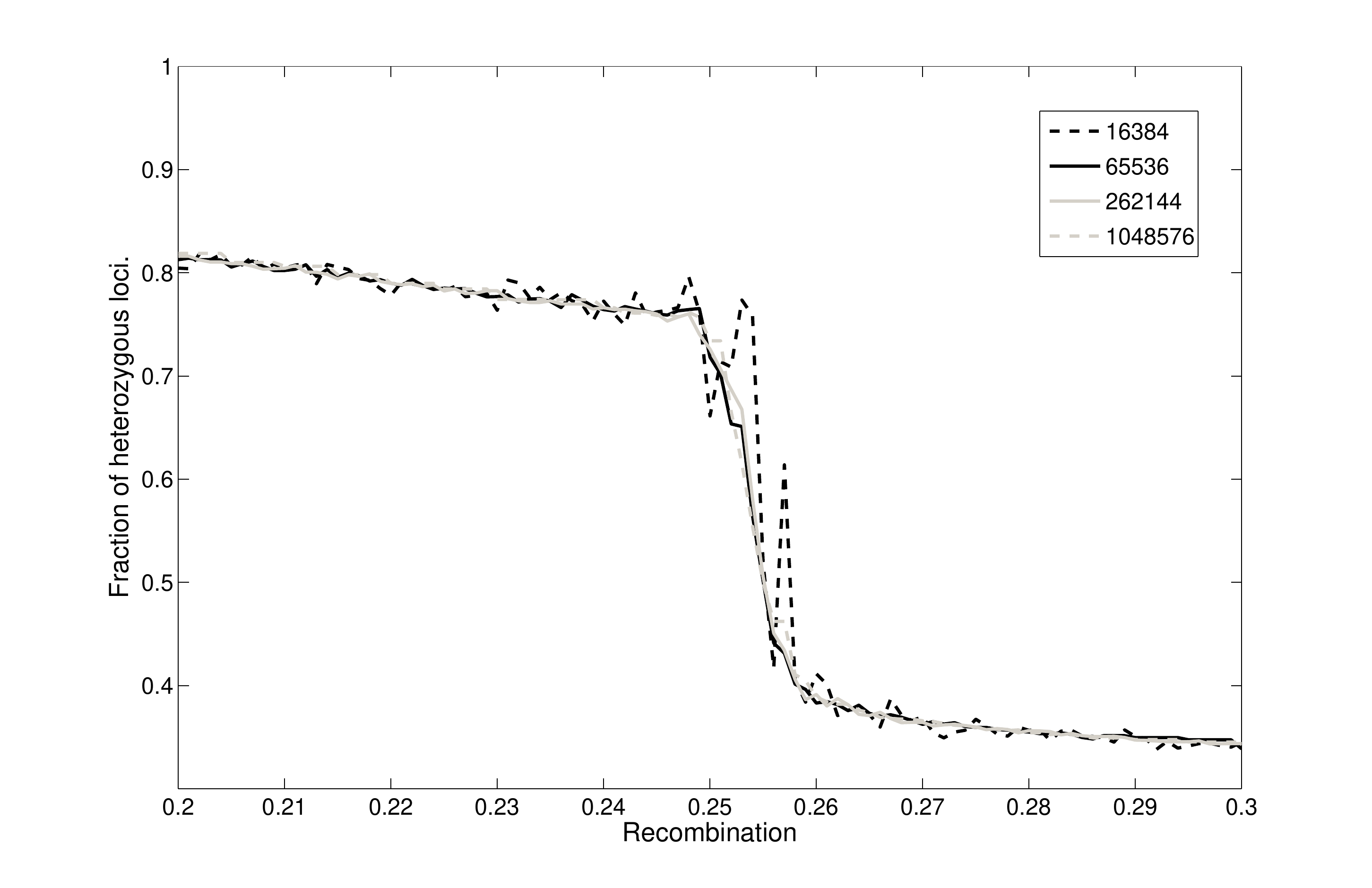}
\caption{Transition between complementing strategy and purifying selection in simulations on square lattice of different sizes (see legend in upper right corner). More detailed information in the text.}
\label{f9}
\end{figure}

One can say that transition between the purifying and complementing strategies of evolution has not the character of phase transition because it depends on population size. In fact the transition directly depends on the probability (frequency) of recombination in the whole genetic linage separating sexual partners. It is possible to show such relations in simulations performed on a square lattice. If a female can choose a partner at distance no larger than P and she can locate her baby at distance no larger than $O$ then the effective population size and inbreeding coefficient is determined by these two parameters: $P$ and $O$, independently of the population size (dimensions of the lattice). Results of simulations (Fig.\ref{f9}) suggest that under all other parameters fixed, $C^*$ value does not depend on population size \cite{waga} \cite{waga2}. For increasing $P$ and/or $O$, $C^*$ decreases and populations enter the purifying strategy at lower recombination rate (results not shown).

\subsection{Zygotic death close to the critical recombination rate and populations’ extinction}

\begin{figure}
\centering
\includegraphics[angle=0,width=\textwidth]{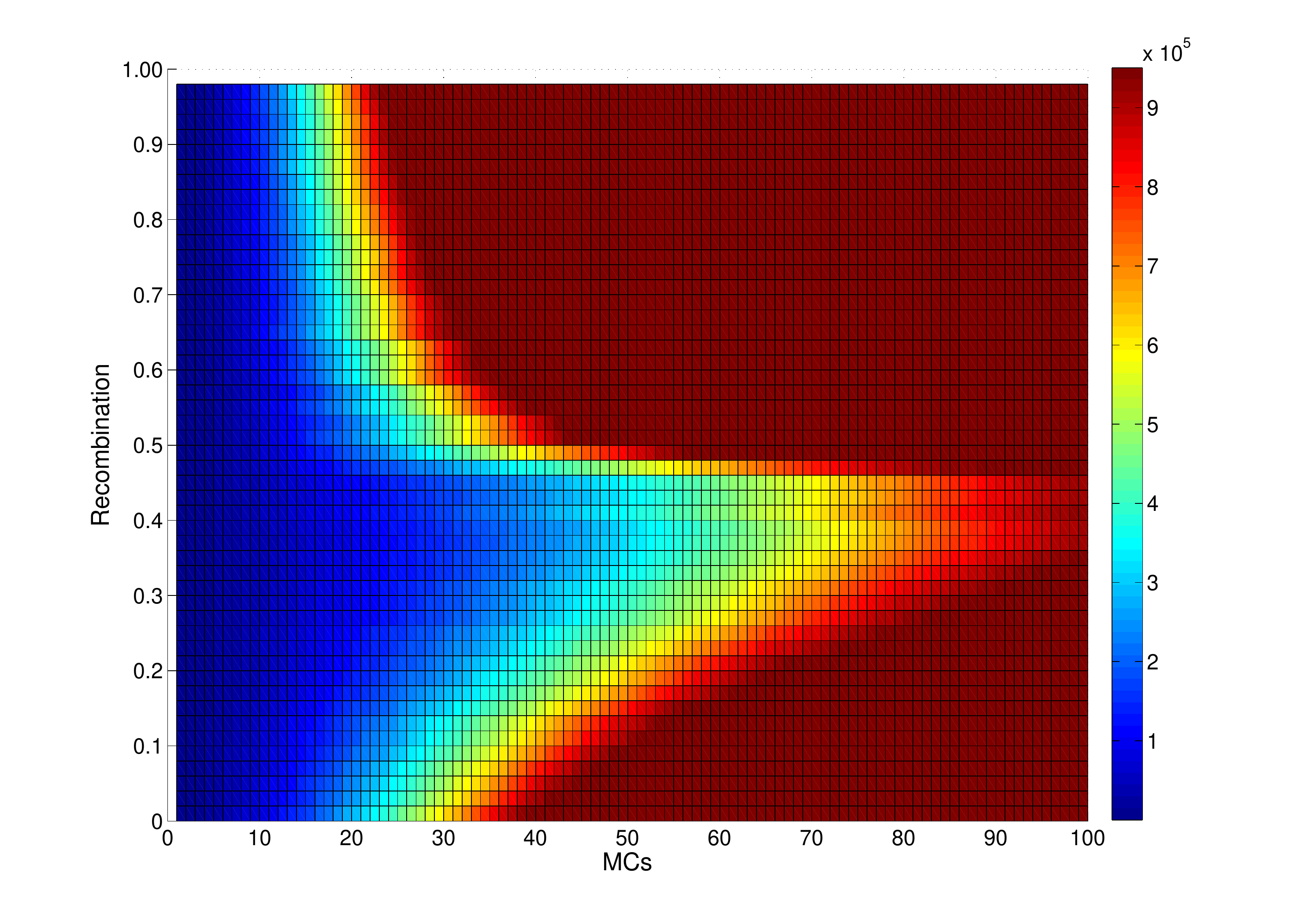}
\caption{Expansion of populations on lattice depending on recombination rate. Blue/red scale shows the number of individuals on lattice after given time of simulations measured in Monte Carlo steps.}
\label{f10}
\end{figure}

In the versions of model where population size was controlled by Verhulst factor instead of being kept constant, it has been observed that population size depends non-monotonically on recombination rate. Under low and high recombination rate populations are larger while in the intermediate recombination rate they are smaller or could be even extinct. This effect is connected with higher zygotic death close to the critical recombination rate (Fig.\ref{f7})\cite{Zawierta2} \cite{waga2}. Zygotic death in this case means, that fusion of two gametes produces a dying offspring. If the birthrate is too low to compensate zygotic death the population is extinct \cite{Stauffer1}. The same effect was observed in populations expanding on square lattices (Fig.\ref{f10}) \cite{waga}. 

\subsection{Relations between chromosome size, coding density and critical recombination frequency}

\begin{figure}
\centering
\includegraphics[angle=0,width=\textwidth]{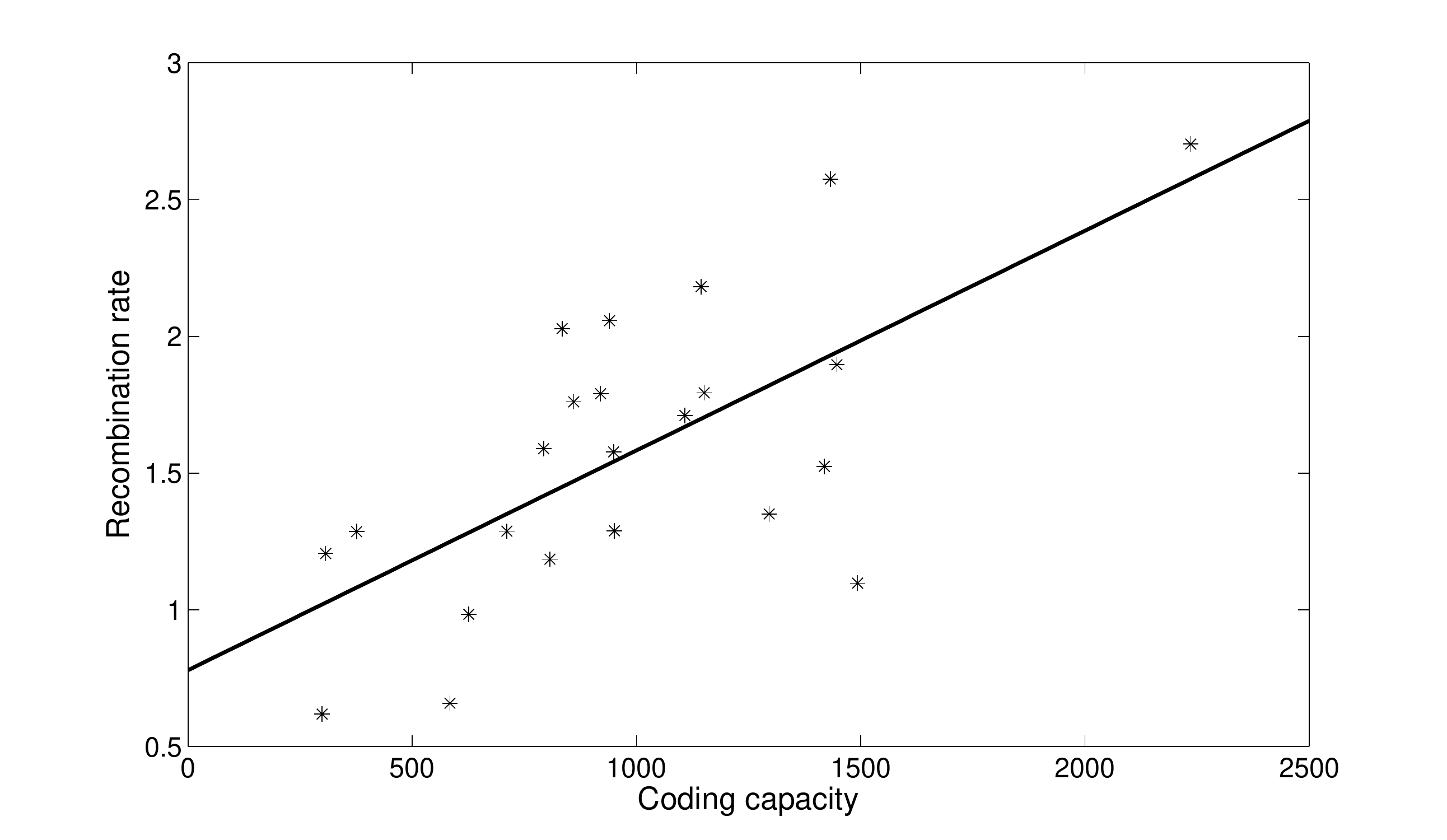}
\caption{Relation between the coding capacity of human chromosomes measured in number of genes (x-axis) and average number of crossover events during one meiosis (y-axis).}
\label{f11}
\end{figure}

\begin{figure}
\centering
\includegraphics[angle=0,width=\textwidth]{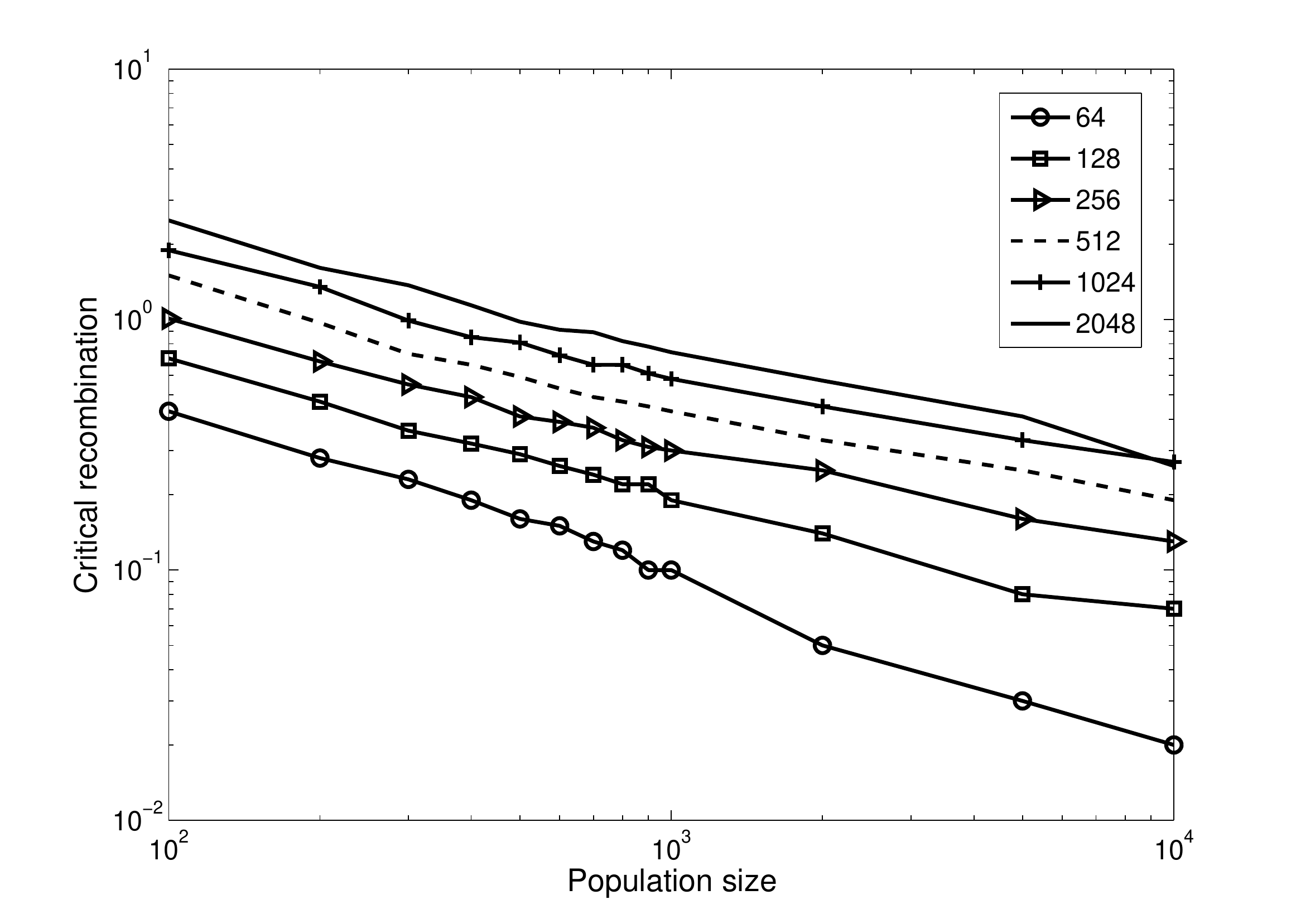}
\caption{Relations between critical recombination rate and effective population sizes for different chromosomes’ lengths (number of bits in the bitstring in the right corner legend). Notice double logarithmic scale.}
\label{f12}
\end{figure}

\begin{figure}
\centering
\includegraphics[angle=0,width=\textwidth]{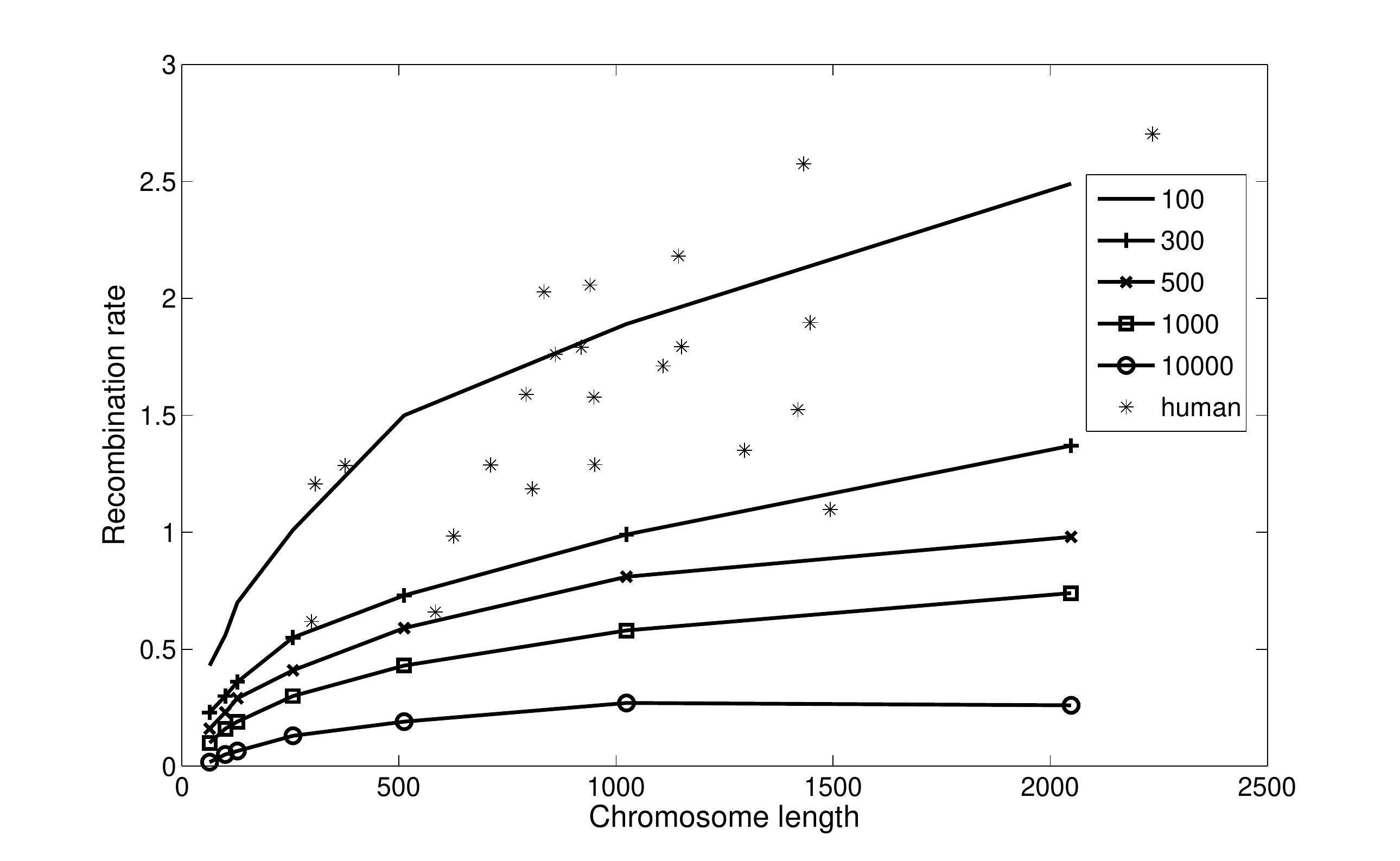}
\caption{Relations between critical recombination rate and chromosome length for different effective population size. Corresponding data for human chromosomes (plotted as stars) suggest that they could be close to critical conditions assuming that humans evolved in populations of effective size of hundreds of people.}
\label{f13}
\end{figure}

The size of natural chromosomes is measured by the number base pairs, by the number of genes located on them or by the genetic units which correspond to the frequency of recombination. One unit (centiMorgan - $cM$) corresponds to the distance between two markers where the probability of recombination during one meiosis is $0.01$. The smallest human chromosome (excluding $Y$ chromosome) contains $280$ genes while the largest one about $2200$ genes. On the other hand genetic length of the shortest one is $62 cM$ while of the largest chromosome is $278 cM$ (for details see International HapMap Consortium 2005 and \cite{Mackiewicz2010}). Since the correlation between coding capacity and recombination rate is not very high in the human genome (Fig.\ref{f11}) - our natural chromosomes may differ significantly in relations between recombination rate and number of genes. The most densely packed human chromosome (19) possesses almost $14$ genes per cM while the least packed one (18) has only $2.5$ genes per $cM$ on average. What is more interesting, these two chromosomes have almost the same length measured in $cM$. In computer modelling usually the number of genes (bits) is considered as a parameter describing a size of chromosome. Recombination is considered as an independent parameter of simulations. In fact there is a very strong relation between these two parameters. In the previously described results it has been shown that chromosomes containing the same number of genes at recombination rate below the $C^*$ value enter the complementing strategy while above the $C^*$ value - the purifying selection. That means that for higher coding density per recombination unit complementation should be expected \cite{Zawierta2} \cite{Mackiewicz2010}. The relations are much more complicated. Generally, there is a power low relation between the $C^*$ value and the effective population size for any chromosome size (Fig.\ref{f12}, notice $log/log$ scales). Nevertheless, increases of $C^*$ values in the $log/log$ plots are observed for growing numbers of genes per chromosome - chromosomes with higher content of genes can enter the complementation strategy under higher recombination rate. Though, the relation is not linear. In Fig.\ref{f13} the relations between $C^*$ and the number of genes per chromosomes are shown for different population size. At the same figure, the data for real human chromosome are plotted (circles). These data suggest that most of human chromosomes could enter complementing strategy if the effective population size is in-between $100$ and $300$. Chromosome $22$ could enter the complementing strategy even if effective population size is of the order of $500$. Computer simulations of evolution of genomes consisting of two pairs of chromosomes differing in the number of genes have shown that one pair of chromosomes can be under the complementing strategy while the other pair under purifying selection. These results were obtained for populations without age structure thus; the effective population size in the modelled populations should be multiplied by the factor of three to correspond the natural effective populations since only about $1/3$ of the population is in reproductive age. The same fraction of individuals being in reproduction age is observed in human populations \cite{Cavalli}. Genetic analyses suggest that humans evolved in the effective populations of this order \cite{Takahata} \cite{Liu}.

\subsection{Complementarity in polymorphic loci}

In all simulations described above only two forms of alleles were considered: functional one - dominant and defective one - recessive. A homozygous locus with both alleles recessive determined the deleterious phenotypic trait. In the nature, relations between alleles are more complicated that is why some simulations have been performed with polymorphic loci. Each gene composed of $8$ bits could be represented by $256$ different alleles, with their adaptive values from $0$ to $8$ depending on the number of bits set to $1$. Thus, the sum of two alleles values occupying the same locus in the homologous chromosomes could vary from $0$ to $16$ and the environment could require a declared value of a given locus. This demand could be fulfilled by different pairs of ''complementing'' alleles. In such simulations, the emerging of complementarity was also observed \cite{Cebrat2009}.

\subsection{Consequences of complementarity}

Computer simulations have shown that critical parameters for emergence of complementarity are recombination frequency, effective population size and coding density of chromosomes. The history of evolution of our species and human genome parameters suggest that the consequences of complementarity should be seen at least in some regions of human chromosomes. Inside the densely packed chromosomes, like $19$, some tendency to grouping genes into clusters is observed \cite{Dehal}. Positions of these clusters should correlate with so-called recombination deserts. Recombination hot spots and deserts seem to be non-randomly distributed on human chromosomes \cite{Jeffreys} \cite{Petes} \cite{Mezard} \cite{Yu}. Is it possible to predict their appearance on the basis of computer simulation results?

\subsection{Distribution of recombination events along chromosomes and gene clustering}

\begin{figure}
\centering
\includegraphics[angle=0,width=\textwidth]{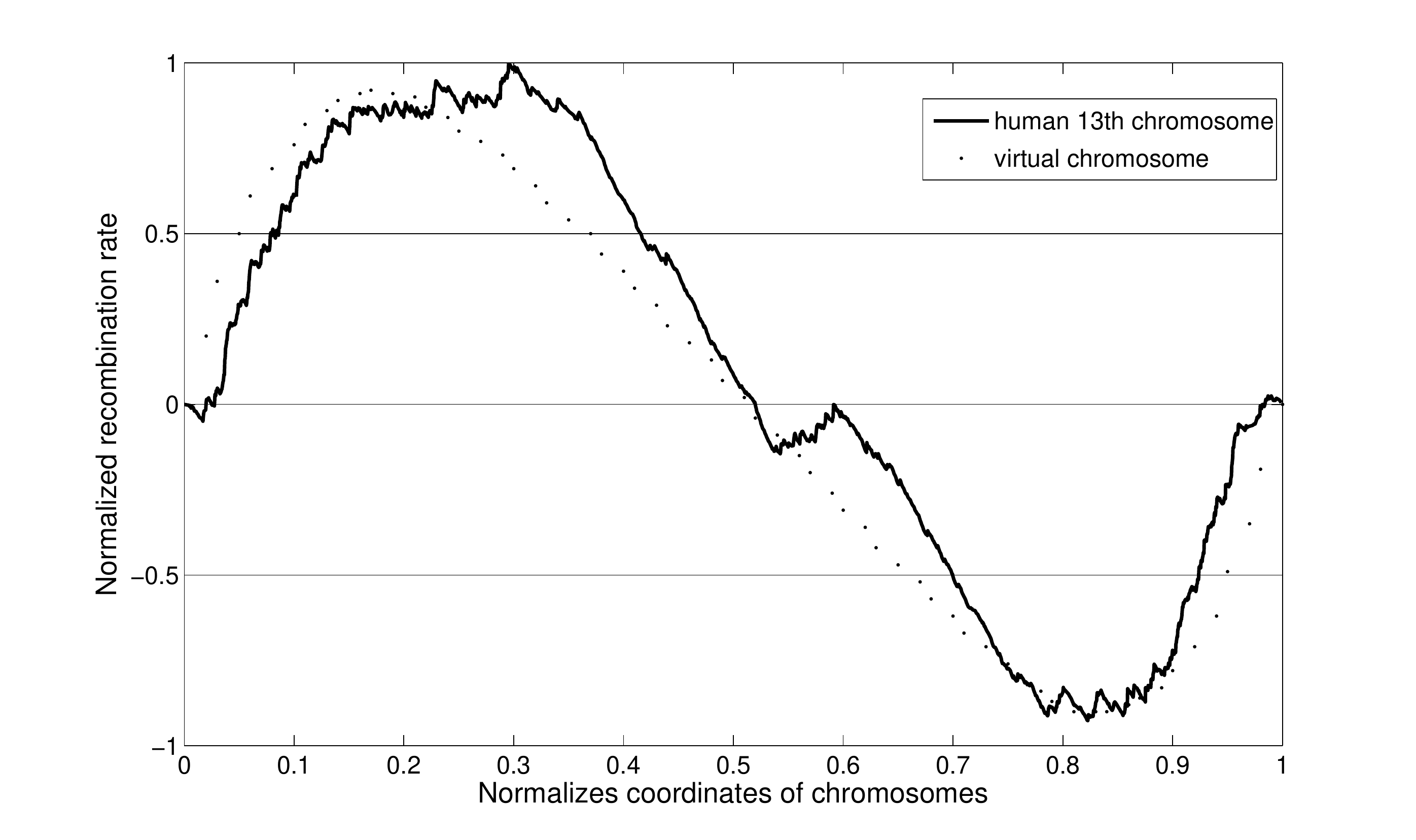}
\caption{Detrended cumulative plots of recombination rate distribution for human $13$ chromosome and virtual chromosome. Increasing regions (subtelomeric) indicate higher than average recombination rate while decreasing region (central part) shows the lower than average recombination rate.}
\label{f14}
\end{figure}

\begin{figure}
\centering
\includegraphics[angle=0,width=\textwidth]{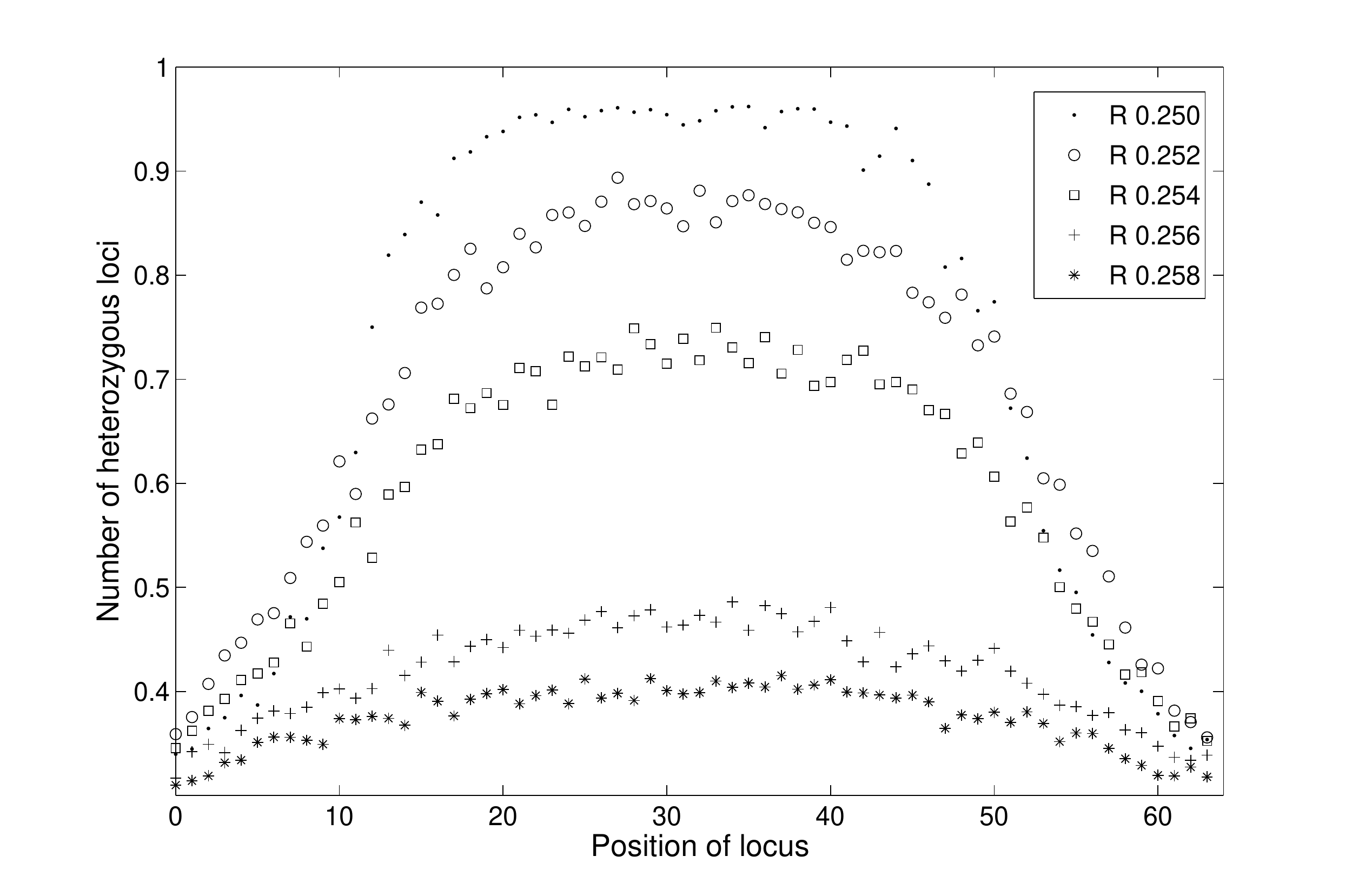}
\caption{Distribution of heterozygous loci along the virtual chromosomes for recombination rate close to critical one. Compare with plots in Fig. \ref{f14}. In the central part of chromosome where recombination rate is relatively low, the fraction of heterozygous loci (complementation) is high.}
\label{f15}
\end{figure}

In all simulations of the role of recombination frequency in the genome evolution, it has been declared that the recombination events are distributed evenly. In the course of simulation, when the distribution of recombinations were analysed, only those events were counted which lead to the gametes producing surviving zygotes. Haplotypes of these gametes were represented in the analysed individuals. We called them ''accepted recombination events''. Usually, when analysing recombinations in the living systems, also only those accepted events are counted. It has been noticed that the distribution of accepted recombination events in the genomes of simulated populations depends on parameters of simulations. If evolution is studied in small effective populations under relatively low recombination rate, the central parts of chromosomes start to form clusters of genes where recombinations have deleterious effect on reproduction potential \cite{waga2}. Gametes which are produced by recombinations in these regions have lower chance to produce surviving zygotes. As a result, the recombination events in gametes which succeed in forming the surviving individuals have a characteristic distribution - a higher recombination frequency in the regions close to the ends of chromosomes (subtelomeric regions) and a lower recombination rate in the central parts of chromosomes \cite{kosmos}. This distribution resembles the distributions observed for example in all human chromosomes, except of $Y$ chromosome \cite{Cheung} \cite{Jensen} \cite{Kong} \cite{Nachman} \cite{Payseur} \cite{Phillips}.  In Fig.\ref{f14}, the distributions of recombinations along the virtual and real chromosomes are plotted. As has been mentioned before, the effects shown in Fig.\ref{f14} apply for accepted recombinations. If one considers the problem from the point of evolutionary costs, it seems to be more effective to avoid the recombination in some regions rather than to eliminate recombinants from the population. It has been shown in simulations that the distributions of recombination spots are under selection pressure and can evolve \cite{waga2} \cite{Kowalski}. To find this, simulations start with recombinations distributed evenly. Frequencies of recombinations at given positions are inherited, though they can be slightly changed from one generation to the other. As a result, the distribution of recombination changes into the characteristic pattern - higher recombination rate in the subtelomeric regions and lower in the central part, simultaneously in the central parts clusters of complementing genes emerge which can be observed as growing fraction of heterozygous loci in the simulated genomes \cite{Zawierta2} (Fig.\ref{f15}) or high linkage disequilibrium in natural chromosomes \cite{Phillips}. There is one very important implication of such an evolution of chromosomal structure - the distribution of recombination events seems to be optimized during evolution. Nevertheless, there are at least two equivalent parameters shaping this structure: recombination frequency and effective population size. It has been shown that the recombination frequency and its distribution along chromosomes could be an effect of long-term genome evolution while the effective population size can change much faster and could be even considered as an environmental condition. Thus, one can expect that an optimum in the effective population size exists for the reproduction potential. For larger and smaller effective populations the reproduction potential should decrease. In fact, it has been found that such an optimum exists for human population \cite{Helgason}. This phenomenon cannot be explained assuming the properties of Mendelian populations. 

\subsection{Sympatric speciation}
\begin{figure}
\centering
\includegraphics[angle=0,width=\textwidth]{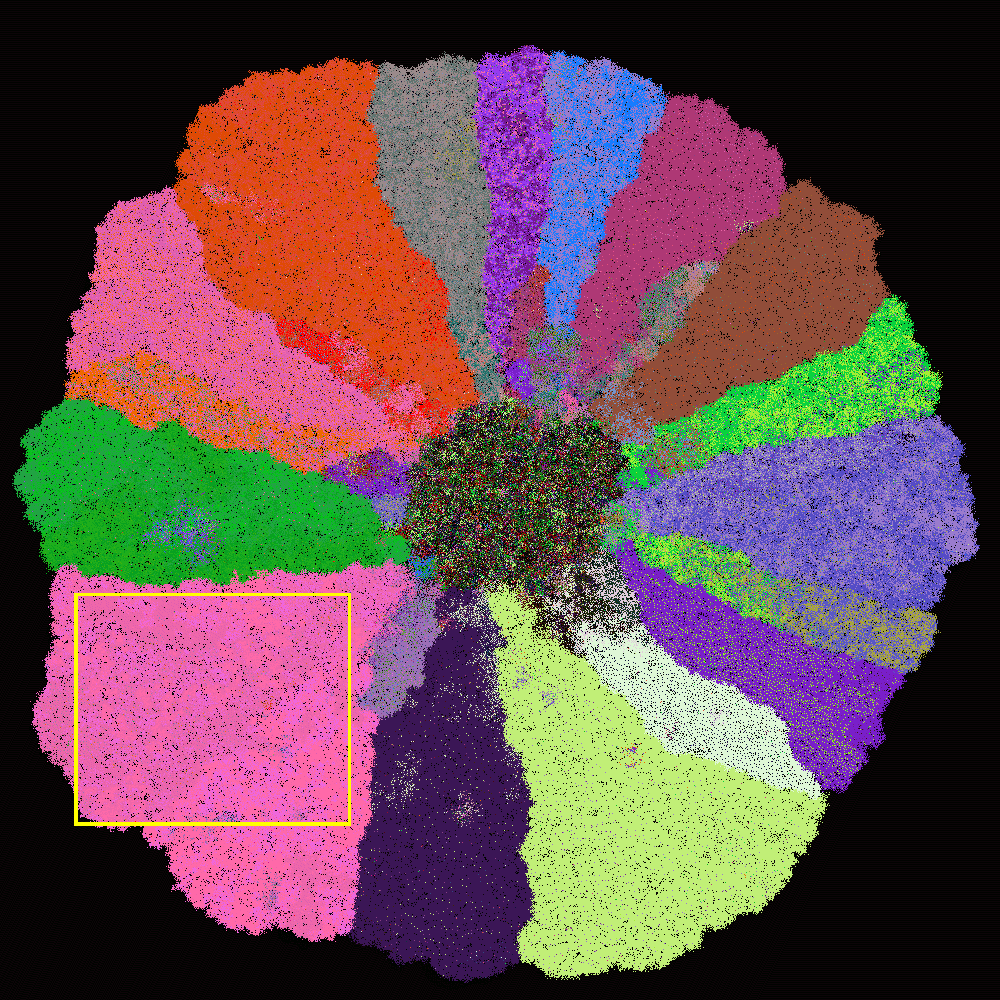}
\caption{Speciation effect during the expansion of population on a square lattice. Population starts from one ideal pair of ancestors (Adam and Eve). Regions of different colours are occupied by different species. They are allowed to hybridize in the border regions but hybrids can not survive because of genetic reasons (haplotypes are not complementing).}
\label{f16}
\end{figure}

\begin{figure}
\centering
\includegraphics[angle=0,width=\textwidth]{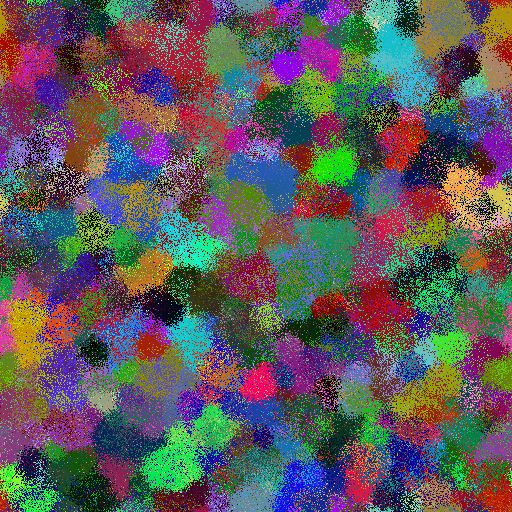}
\caption{Sympatric speciation on square lattice wrapped on torus. Initial population was randomly generated (age and genetic structure). The rest parameters like in Fig. \ref{f16}.}
\label{f17}
\end{figure}

\begin{figure}
\centering
\includegraphics[angle=0,width=\textwidth]{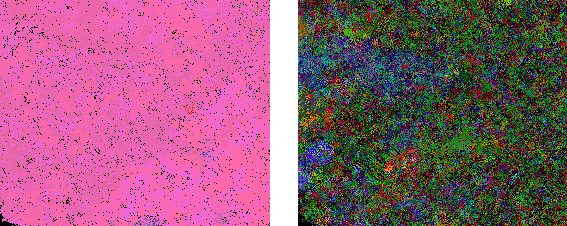}
\caption{Fragment of population marked in Fig. \ref{f16}. Left panel coloured according to central part of chromosomes, right panel coloured according to lateral parts of chromosomes. Central parts seems to be more homogenous, conservative responsible for speciation effect and characteristic for a given species while the lateral parts are more polymorphic responsible for different individual traits inside species.}
\label{f18}
\end{figure}

Sympatric speciation is a phenomenon of a new species emerging inside the population of an existing species without any physical, geographical or biological barriers. For a long time it has been considered as a negligible phenomenon in the biological evolution \cite{Mayr}. Recently it is still debatable \cite{Jiggins} though, there are more and more known examples of such an evolution and sympatric speciation attracts more attention \cite{Cichlids} \cite{Bunje}. In the large Mendelian, homozygous populations sympatric speciation seems to be really impossible but if one considers a spatial distribution of individuals, simple physical distances between them can introduce some restrictions into the reproduction process - i.e. higher probability of finding a sexual partner in the neighbourhood rather than at long distances. Simulations on lattices with declared maximum distances for looking for partners and locating the offspring enable the emergence of heterogeneity of populations. In such populations the genetic relations between individuals are negatively correlated with distances between them. On the other hand the probability of mating between two individuals decreases with the distance. It is possible to follow (or almost directly observe) the evolution of individual genomes on lattice \cite{waga} \cite{waga2}. Let assume that to each number from the range $1$ to $2^{24}$ a specific colour is ascribed and a $24$ bits long fragment of a bitstring, representing one haplotype of a diploid genome is considered as a number in a binary numeral system. Thus, a place when an individual is located is coloured according to $24$ bits fragment of its genome. Fig.\ref{f16} illustrates the evolution of population which started from one pair of ideal individuals and expanded onto the whole lattice. Colours of large radiating regions show the territories occupied by individuals with the same fragment of haplotype determining the colour. In Fig.\ref{f17} another case is presented. Lattice was wrapped on torus and the initial population was composed of individuals with randomly set genomes and age. In the course of simulation the whole initial population was divided for subpopulations seen as distinctly coloured territories. In both cases, individuals located close to the borders of those distinctly coloured regions can interbreed and form hybrids but these hybrids can not survive up to their reproduction age - by definition, the individuals occupying those differently coloured territories belong to different species. In both cases it was complementation which was responsible for speciation. Relatively low recombination rate and high inbreeding established by distances $P$ and $O$ generate the complementing clusters of genes. Since such a complementation starts in different parts of lattice independently, the structure of complementing clusters are different and do not fit to each other. Thus, emerging subpopulations with different clusters behave like the new different species. 
Mechanisms of complementing strategies described in previous sections show that complementation starts in the central parts of chromosomes. Lateral fragments could be under purifying selection. It could be also directly observed in the results of simulation. The subpopulation inside the square shown in Fig.\ref{f16} seems to be very homogenous according to the central fragment of individuals' genomes (Fig.\ref{f18} panel $A$). But if the same fragment of population is coloured according to the lateral parts of chromosomes, then the populations seems to be very polymorphic (Fig.\ref{f18} panel $B$). According to this observation, central parts of chromosomes should be rather conservative and responsible for belonging to higher order taxons while in the lateral parts, less conservative genes should be located. In the real human chromosomes, the fraction of genes which have their homologous counterparts in the mouse genome is lower in subtelomeric regions than in the central parts \cite{kosmos}.
There is another interesting feature of sympatric speciation observed in virtual populations on lattices. As it has been noted, speciation is facilitated when intragenomic recombination is low and inbreeding is high (small effective population). Intragenomic recombination rate is fixed by evolution and probably changes slowly as an important feature of natural genomes. Inbreeding can change faster even following the environmental changes. If the fatal environmental disaster reduces the population on a given territory, the inbreeding is growing and speciation can occur like during the Permian. It has been also shown that sympatric speciation is faster in a changing environment even if the whole environment is changed uniformly \cite{Waga2009}.

\subsection{Gamete recognition}
Complementation strategy assumes that two different (complementing) sequences of alleles fit to each other producing a better fitted diploid genome. Assuming that defective alleles have recessive lethal characters, zygotes containing two identical clusters of genes should die and only those containing complementing clusters would survive. If we consider a set of chromosomes with only one pair of complementing clusters then the probability of zygotic death is $0.5$. Evolutionary costs and reproduction would be more efficient if gametes could recognize which of them has an identical cluster and which one has a complementing cluster of genes before fusing to form a zygote \cite{Cebrat2008}. Such systems of recognition or probing the information inside another cell are known even in prokaryotes - i.e. an entry exclusion system which prevents bacteria to engage in conjugational process if a partner cell already possesses genetic information to be transferred \cite{Morzejko} \cite{Novick}. It is suggested that in humans the \emph{Major Histocompatibility Complex (MHC)} can play such a role in preselection of mating partners or in the very early immune reaction of mother on embryo antigens \cite{Jacob} \cite{Parham}. The \emph{MHC} complex alone is not enough to guarantee the fusion of complementing haplotypes. If complementing clusters can appear on many chromosome pairs - all these pairs should possess an operating recognition system. The mechanism should be located at the level of gametes and, to be efficient, it should be independent for different pairs of chromosomes nestling the complementing clusters of genes. There is a group of genes which could fulfil such a role - \emph{Olfactory Receptor genes (OR)} \cite{Vosshall}. This is the largest gene family in the human genome composed of almost $1000$ genes, clustered in many different groups located on almost all chromosomes (excluding $20$ and $Y$) and at least some of these genes are expressed during spermatogenesis \cite{Parmentier} \cite{Goto}. Thus, OR could play a role in chemotaxis of sperm cells enhancing the chance of the best fitted one for fertilization the egg \cite{Spehr}. If we assume that each of our $22$ pairs of autosomes have complementing clusters of genes, then a recognizing system should increase the probability of successful fertilization by several orders of magnitude. On the other hand, the existence of such a system explains why a male produces such an enormous surplus of sperm cells.

\bibliography{review}{}
\bibliographystyle{plain}

\end{document}